\documentclass[%
aip,jcp,
sd,%
amsmath,amssymb,
reprint,%
]{revtex4-1}

\usepackage{graphicx}             
\usepackage{dcolumn}                
\usepackage{bm}                          
\usepackage[mathlines]{lineno} 

\usepackage{natbib}
\usepackage{fleqn}
\usepackage{hyperref}
\usepackage{verbatim}

\usepackage{todonotes}

\begin{document}


\title{Successes and failures of Hubbard-corrected density functional theory: The case of Mg doped LiCoO$_2$}

\author{Juan A. Santana}
\affiliation{Materials Science and Technology Division, Oak Ridge National Laboratory, Oak Ridge, TN 37831, USA}

\author{Jeongnim Kim}
\affiliation{Materials Science and Technology Division, Oak Ridge National Laboratory, Oak Ridge, TN 37831, USA}

\author{P. R. C. Kent}
\affiliation{Center for Nanophase Materials Sciences, Oak Ridge National Laboratory, Oak Ridge, TN 37831, USA}
\affiliation{Computer Science and Mathematics Division, Oak Ridge National Laboratory, Oak Ridge, TN 37831, USA}

\author{Fernando A. Reboredo}
\email{reboredofa@ornl.gov}
\affiliation{Materials Science and Technology Division, Oak Ridge National Laboratory, Oak Ridge, TN 37831, USA}

\date{\today}

\begin{abstract}
 
We have evaluated the successes and failures of the Hubbard-corrected density functional theory (DFT+U) approach to study Mg doping of  LiCoO$_2$. We computed the effect of the U parameter on the energetic, geometric and electronic properties of two possible doping mechanisms: (1) substitution of Mg onto a Co (or Li) site with an associated impurity state and, (2) formation of impurity-state-free complexes of substitutional Mg and point defects in LiCoO$_2$. We find that formation of impurity states results in changes on the valency of Co in LiCoO$_2$. Variation of the Co U shifts the energy of the impurity state, resulting in energetic, geometric and electronic properties that depend significantly on the specific value of U. In contrast, the properties of the impurity-state-free complexes are insensitive to U. These results identify reasons for the strong dependence on the doping properties on the chosen value of U and for the overall difficulty of achieving agreement with the experimentally known energetic and electronic properties of doped transition metal oxides such as LiCoO$_2$.

\end{abstract}


\maketitle

\section{Introduction}

Although theory has been used to support research in new rechargeable battery materials,\cite{ceder_recharging_2011} whether theory can be used to design new materials without experimental input will largely depend on the ability of electronic structure methods to describe new and existing materials with accuracies compatible with experimental needs.\cite{kolorenc_applications_2011,booth_towards_2013,shulenburger_quantum_2013}  Thus, in order to respond to initiatives such as the Materials Genome,\cite{comment1} the predictive power of theory must be characterized.
 
In energy storage research, a significant  challenge is to accurately predict the thermodynamic phase stability and ionic/electronic conductivity of cathode materials.\cite{ceder_recharging_2011,wolverton_first-principles_1998,van_der_ven_first-principles_1998,marianetti_first-order_2004} This is because theoretical calculations are mainly based on Density Functional Theory (DFT) within the local density (LDA) and generalized gradient (GGA) approximations. While these approximations are accurate enough to study the thermodynamic and conductive properties of many materials, they are unreliable in others, e.g., transition metal-oxides. Approximations such as LDA and GGA do not account properly for exchange and correlation effects in transition metal-oxides, leading to exchange-correlation errors such as the self-interaction error.\cite{perdew_self-interaction_1981} In practice a semi-empirical approach is adopted in an attempt to minimize these errors, but the use of experimental data for the calibration and validation inevitably reduces the predictive power of the method.

The doping of cathode materials is an example where it is crucial to accurately predict thermodynamic and conductive properties and where a predictive theory is highly desired. A case that exemplifies the problem is the doping of LiCoO$_2$ with cations such as Mg. The effect of Mg-doping on the electronic conductivity and long-term capacity retention of layered LiCoO$_2$-based  batteries have been extensively studied.\cite{carewska_electrical_1997,tukamoto_electronic_1997,julien_combustion_2000,mladenov_effect_2001,julien_structural_2002,levasseur_dual_2002,madhavi_cathodic_2002,chen_electronic_2003,xu_electronic_2003,frangini_voltammetric_2003,li_enhancement_2003,elumalai_microwave_2004,kim_electrochemical_2004,nobili_ac_2005,xu_improving_2005,kim_crystal_2006,sathiyamoorthi_layered_2007,eom_m3_2008,zaheena_microwave_2009,zhou_relative_2009,luo_synthesis_2010,valanarasu_effect_2010,jung_improved_2012,nobili_solgel_2012}
Carewska \textit{et al.}\cite{carewska_electrical_1997} and Tukamoto
and West\cite{tukamoto_electronic_1997} showed that doping LiCoO$_2$ with Mg increases its electronic conductivity. This higher conductivity was rationalized by considering the generation of electronic holes due to the formation of a mixed +3/+4 valence state in Co.\cite{carewska_electrical_1997,tukamoto_electronic_1997} Later studies\cite{chen_electronic_2003,xu_electronic_2003,xu_ab_2004,xu_ab_2005,shi_effect_2007} supported this model and suggested that the transfer of an electron from Co ion to O 2\textit{p} hole is also involved in raising the conductivity. The formation of oxygen vacancies in Mg-doped LiCoO$_{2}$ has also been proposed to contribute to the higher conductivity.\cite{levasseur_dual_2002,madhavi_cathodic_2002,luo_synthesis_2010} 

In addition to the increased conductivity, a significant capacity retention for Mg-doped LiNiCoO$_2$-based batteries was independently reported by Chang \textit{et al.},\cite{chang_synthesis_2000} Cho\cite{cho_lini0.74co0.26-xmgxo2_2000} and Kweon \textit{et al.}\cite{kweon_modification_2000} For example, Cho found 92\% capacity retention for LiNi$_{0.74}$Co$_{0.22}$Mg$_{0.04}$O$_2$ after 94 cycles at 1 C rate vs. 70\% for LiNi$_{0.74}$Co$_{0.26}$O$_2$.\cite{cho_lini0.74co0.26-xmgxo2_2000} Mg-doping also improves the thermal stability of  Li-ion batteries. Similar improved capacity and thermal properties were later reported by other authors for LiCoO$_2$-based batteries. \cite{julien_combustion_2000,mladenov_effect_2001,julien_structural_2002,levasseur_dual_2002,chen_electronic_2003,frangini_voltammetric_2003,li_enhancement_2003,xu_electronic_2003,elumalai_microwave_2004,kim_electrochemical_2004,nobili_ac_2005,xu_improving_2005,kim_crystal_2006,sathiyamoorthi_layered_2007,eom_m3_2008,zaheena_microwave_2009,zhou_relative_2009,luo_synthesis_2010,valanarasu_effect_2010,jung_improved_2012,nobili_solgel_2012} However, doping LiCoO$_{2}$ and LiNiO$_{2}$ based cathode materials with Mg cations has the drawback of decreasing the capacity\cite{tukamoto_electronic_1997,chang_synthesis_2000,cho_lini0.74co0.26-xmgxo2_2000,kweon_modification_2000,mladenov_effect_2001} because Mg reduces the concentration of the active Ni$^{3+}$ or Co$^{3+}$ ion sites.\cite{tukamoto_electronic_1997,chang_synthesis_2000,cho_lini0.74co0.26-xmgxo2_2000}

Several explanations for the stability of Mg-doped Li-ion batteries have been proposed. Cho proposed that a lower cation mixing in LiNi$_{0.74}$Co$_{0.26-x}$Mg$_{x}$O$_2$ than in LiNi$_{0.74}$Co$_{0.26}$O$_2$ battery leads to the improved stability.\cite{cho_lini0.74co0.26-xmgxo2_2000} A large degree of cation mixing, where Li ions partially occupy the Ni (Co) sites and vise versa, results in lower capacity and affects the structural stability of the layered material.\cite{cho_lini0.74co0.26-xmgxo2_2000,cho_effect_1999}

The higher stability has also been rationalized as a pillaring effect, where Mg located in the oxide layers or inter-layer spaces prevents the structural collapse and crystallinity loss during charge/discharge. However, it is still unclear if Mg is located in oxide layers or inter-layers spaces, i.e. whether Mg is on Li or transition metal sites. It has been proposed that Mg is initially on the transition metal sites, but it migrates to the inter-layer spaces after initial cycling.\cite{pouillerie_synthesis_2000,pouillerie_effect_2001,kim_electrochemical_2004,depifanio_effect_2004}
Xiang \textit{et al.}\cite{xiang_rheological_2008} proposed a similar model based on a detailed analysis of measured change in the volume of LiNi$_{0.80-x}$Co$_{0.20}$Mg$_{x}$O$_2$ as function of \textit{x} and possible substitutions and charge balance mechanisms for Mg. They suggested that the stronger Mg-O bond (vs. Li-O) enhances the stability of LiNi$_{1-x}$Co$_x$O$_2$-based batteries. On the other hand, Chang \textit{et al.}\cite{chang_synthesis_2000} found that Mg cations are mainly on Ni sites. They argued that Mg stabilizes the NiO$_2$ slab, preventing thermal and cycling decomposition.\cite{chang_synthesis_2000} Tatsumi \textit{et al.}\cite{tatsumi_local_2008} also found Mg cations to preferentially replace Ni sites initially, but to diffuse out of the active material during cycling. More recently, Tavakoli \textit{et al.}\cite{tavakoli_stabilizing_2013} proposed that a short-range ordering of Ni cations around Mg results on LiNi$_{0.755}$Co$_{0.147}$Al$_{0.045}$Mg$_{0.053}$O$_2$ batteries 34.2 $\pm$ 9.3 meV more thermodynamically stable than undoped LiNi$_{0.800}$Co$_{0.155}$Al$_{0.045}$O$_2$. Such stabilization can provide stronger bonding and prevent the formation of NiO-like phase during charge/discharge.\cite{tavakoli_stabilizing_2013}

Despite the significant experimental efforts outlined above, it is still unclear how Mg-doping improves the stability of LiCoO$_2$ and LiNiO$_2$ based batteries. In part, this is because no theoretical work has been performed to evaluate the thermodynamic profile of Mg in these batteries. Calculations have mainly focused on the electronic properties,\cite{xu_electronic_2003,xu_ab_2004,xu_ab_2005} lattice stability\cite{shi_effect_2007}  and Li-intercalation voltage\cite{shi_effect_2007} of Mg-doped LiCoO$_2$. Calculations have also been employed to study the site preference of Mg in LiNiO$_2$ by combining theory, x-ray absorption near-edge structure and electron energy-loss near-edge structure measurements.\cite{tatsumi_local_2008} 

The relative lack of calculations for the thermodynamics of Mg in these cathode materials comes because such calculations of impurities in transition metal-oxides are rather difficult within the framework of DFT. This is mainly because of DFT errors in transition metal-oxides. These errors can be partially removed by introducing an on-site Hubbard model correction (DFT+U).\cite{anisimov_first-principles_1997} This method has been successfully used to study Li-ion battery cathodes,\cite{zhou_first-principles_2004,wang_first-principles_2007,tatsumi_local_2008,kramer_tailoring_2009,koyama_defect_2012} and other transition metal-oxides.\cite{wang_oxidation_2006, jain_formation_2011, jain_commentary:_2013} It is also the standard approach adopted by the ``Materials Project''\cite{jain_commentary:_2013} for high throughput computation of materials, including metal-oxides. However, physically, the parameter U should depend on the chemical environment of the atomic site where it is applied, while in conventional DFT+U calculations a universal value of U is used. As a result DFT+U often fails to correctly predict the relative energy between systems with a mixture of localized and delocalized electronic states.\cite{jain_formation_2011} This is particularly problematic if one uses DFT+U to study dopants that can induce localized or delocalized states in transition metal-oxides.

Mg is a divalent dopant and its substitution onto a Co or Li site can induce a localized or delocalized state in LiCoO$_2$. Therefore, to elucidate how Mg improves the performance of the LiCoO$_2$ cathode material using DFT/DFT+U methods, a first step is to study how sensitive the calculated properties of Mg-doped LiCoO$_2$ are to different choices of the parameter U. To this end, we studied the effect of U on the energetic, geometric and electronic properties of Mg when it is located on Co and Li sites as well as when it forms complexes with Li vacancies or interstitial sites in LiCoO$_2$.

The remainder of this article is organized as follows. We first discuss the doping mechanisms that were examined, the computational methods employed and how the chemical potentials were established. We then present our results. We start by analyzing the effect of the U parameter on the range of chemical potentials, emphasizing the effect on the phase boundaries of LiCoO$_2$. Subsequently, we discuss the effect of U on the formation energy, geometry and electronic structure of Mg-doped LiCoO$_2$. We conclude with a summary and our conclusions.

\section{Methodology}

\subsection{Substitutional Mg in LiCoO$_2$}

Mg is formally a divalent dopant, Mg$^{2+}$, and its substitution onto a Co or Li site in LiCoO$_{2}$ is an aliovalent substitution that requires a mechanism for charge compensation. One possible mechanism is the formation of new electronic species. In this mechanism, the substitution of Mg$^{2+}$ onto a Co$^{3+}$ site leads to the extrinsic Mg$_{Co}$ defect and a concomitant increase in the oxidation state of a Co site from 3+ to 4+, which introduces an impurity hole in the system. This mechanism was originally proposed to rationalize the high conductivity of Mg-doped LiCoO$_{2}$.\cite{carewska_electrical_1997, tukamoto_electronic_1997} On the other hand, the substitution of Mg$^{2+}$ onto a Li$^{+}$ site is balanced by the formation of a Co site with oxidation state +2, i.e an impurity electron is introduced in the system. 

Other mechanisms for charge compensation can involve the formation of vacancy, interstitial and anti-site defects, and defect complexes.\cite{xiang_rheological_2008} In LiCoO$_{2}$, many intrinsic defects can combine with substitutional Mg and form a complex for charge compensation. Mg$_{Co}$ can form a complex with electron donor defects such as V$^{+}_{O}$, Co$^{2+}_{\textit{i}}$, Li$_{\textit{i}}$ and Co$^{+}_{\textit{Li}}$, while Mg$_{Li}$ with electron acceptor defects such as V$_{Li}$, V$^{2-}_{Co}$, Li$^{-}_{\textit{Co}}$. Recent calculations\cite{koyama_defect_2012,hoang_defect_2014} have shown that, of these ionic defects, the dominant one under typical synthesis conditions is the anti-site Co$^{+}_{\textit{Li}}$ defect. The other intrinsic defects have high formation energies and are expected to have a low concentration.\cite{koyama_defect_2012} Based on such reports, we considered in our initial calculations the Mg$_{Co}$-Co$^{+}_{\textit{Li}}$ complex as a charge compensation mechanism. However, the formation energy of this complex is over 1.5 eV when the system is under charge neutrality and, therefore, it was not further considered.\cite{comment2} Instead, we considered the neutral complexes Mg$_{Co}$-Li$_{\textit{i}}$ and Mg$_{Li}$-V$_{Li}$ as they can be relevant to the electrochemistry of Mg-doped LiCoO$_2$. Additionally, we studied the charge compensation by the dual substitution of Mg on Co and Li sites, i.e. the Mg$_{Co}$-Mg$_{Li}$ complex.

To quantify the incorporation of Mg in LiCoO$_2$, we evaluated the formation energy of the possible extrinsic defects and defect complexes as:

\begin{eqnarray}
E^f \left(X\right)=E_{tot} \left[X\right] - E_{tot} \left[LiCoO_2\right]  - \sum_{i}n_i\mu_i
\label{eq:six}
\end{eqnarray}

\noindent where \textit{X} is a neutral extrinsic or intrinsic defect (or defect complex). E$_{tot}$\textit{[X]} and E$_{tot}$\textit{[LiCoO$_{2}$]} are the total energy of LiCoO$_{2}$ containing \textit{X} and the total energy of the equivalent bulk LiCoO$_{2}$, respectively. \textit{n$_i$} is the number of atomic species \textit{i} added (\textit{n$_i$} \textgreater 0) or removed (\textit{n$_i$} \textless 0) from the supercell, while ${\mu}$$_{\textit{i}}$ indicates the corresponding atomic chemical potentials. The stability of the complexes is quantified by their binding energies as:\cite{van_de_walle_first-principles_2004}

\begin{eqnarray}
E_b=E^f \left(X\right) + E^f \left(Y\right) - E^f \left(XY\right)
\label{eq:seven}
\end{eqnarray}

\noindent where E$^{f}$\textit{(XY)}, E$^{f}$\textit{(X)} and E$^{f}$\textit{(Y)} are the formation energy of the complexes and those of the individual defects, respectively. In this notation, a positive binding energy indicates a bound complex. 

\subsection{DFT Calculations}

The total energies to evaluate Eq.~(\ref{eq:six}) were calculated within the DFT framework as implemented on the Vienna Ab-initio Software Package (VASP).\cite{kresse_ab_1993, kresse_ab_1994, kresse_efficiency_1996, kresse_efficient_1996} We used the Perdew-Burke-Ernzerhof (PBE)\cite{perdew_generalized_1996,perdew_generalized_1997} exchange and correlation functionals. The Li, O, Mg and Co ionic cores were represented by the projector augmented-wave (PAW) potentials\cite{blochl_projector_1994,kresse_ultrasoft_1999} with 3, 6, 8 and 9 valence electrons, respectively. The wavefunction energy cutoff was set to 520 eV. We initialized the transition metal atoms in both high and low spin states with ferromagnetic ordering and the configuration with the lowest energy was used. The k-point mesh employed to calculate the bulk properties of LiCoO$_2$, CoO, Co$_3$O$_4$, Li$_6$CoO$_4$, Li$_8$CoO$_6$, Li$_2$O$_2$, Li$_2$O and MgO was 6$\times$6$\times$6, 8$\times$8$\times$8, 4$\times$4$\times$4, 4$\times$4$\times$4, 4$\times$4$\times$2, 6$\times$6$\times$6, 7$\times$7$\times$7, and 8$\times$8$\times$8, respectively. Gaussian broadening with an energy width of 0.1 eV was used for the Brillouin zone integration. These choices are sufficient to converge the bulk energies to better than 2 meV per primitive cell, which is substantially smaller than the variation due to the use of different functionals. 

\begin{figure*}
\includegraphics[scale=0.9]{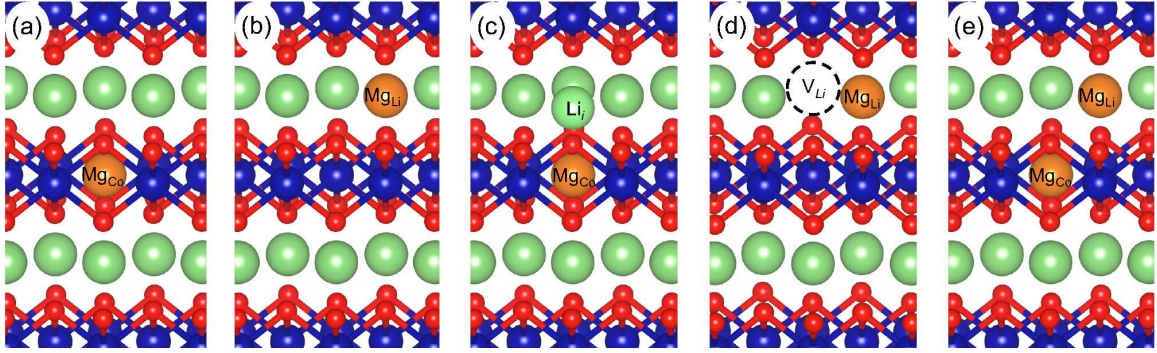}
\caption{\label{fig:fig1} Structures of Mg-doped LiCoO$_2$: a) Mg$_{Co}$, Mg on a Co site; b) Mg$_{Li}$, Mg on a Li site; c) Mg$_{Co}$-Li$_{\textit{i}}$, complex of Mg$_{Co}$ and a Li interstitial; d) Mg$_{Li}$-V$_{Li}$, complex of Mg$_{Li}$ and a Li vacancy; and e) Mg$_{Co}$-Mg$_{Li}$, complex of Mg$_{Co}$ and Mg$_{Li}$. Colored spheres indicate Mg (orange), Li (green), Co (blue), and O (red) atoms, respectively. Image generated with VESTA.\cite{momma_vesta:_2008}}%
\end{figure*}

LiCoO$_{2}$ supercells with a defect \textit{X} were calculated employing 4$\times$4$\times$2 supercells built from the primitive rhombohedral unit cell and  2$\times$2$\times$3 Monkhorst-Pack k-point meshes. The defects and defect complexes that were studied are shown schematically in Fig.~\ref{fig:fig1}. The extrinsic Mg defects were created by the substitution of a Co or Li atom in the Li$_{32}$Co$_{32}$O$_{64}$ supercell. The Li vacancy site was built by removing a Li from the supercell while the Li interstitial-site was constructed by adding a Li atom at the tetrahedral site in the Li layer. To simulate the Mg$_{Co}$-Li$_{\textit{i}}$ and Mg$_{Li}$-V$_{Li}$ defect complexes in our supercell, we evaluated various configurations where the intrinsic Li defects were a first, second or third neighbors of the Mg site. These preliminary calculations indicated that the complexes are more stable when the Li defects are first neighbor to Mg. We found similar results for the Mg$_{Co}$-Mg$_{Li}$ complex. Therefore, the extrinsic Mg defect is always a nearest neighbor of the second defect in the defect complexes that were studied. For each defect, all atomic positions were optimized until residual forces were less than 0.02 eV/$\mathring{\textnormal{A}}$. The volume of the supercell with the defects was fixed to that calculated for the primitive unit cell. 

The use of supercells to study defects introduces quantum mechanical, elastic and electrostatic artifacts.\cite{van_de_walle_first-principles_2004} These artifacts become smaller as the supercell size is increased. In the present calculations, where we evaluated formation energies for neutral defects, only quantum mechanical (wavefunction overlap) and elastic effects are expected. We performed calculations with a larger 4$\times$4$\times$4 supercell to corroborate that the formation energies evaluated with a 4$\times$4$\times$2 supercell are relatively insensitive to these supercell size effects. For the substitution of Mg onto a Co site in LiCoO$_{2}$, the formation energy evaluated with the 4$\times$4$\times$2 supercell is 0.52 eV while with the 4$\times$4$\times$4 cell it is 0.55 eV. For the Mg$_{Co}$-Mg$_{Li}$ complex, the formation energies calculated with the 4$\times$4$\times$2 and 4$\times$4$\times$4 supercells are 0.71 eV and 0.67 eV, respectively. We also evaluated the effect of volume relaxation for the substitution of Mg onto Co with the 4$\times$4$\times$2 supercell; the formation energy is 0.51 eV when volume relaxation is allowed, a 0.01eV difference from the fixed volume result.

The formation energies and electronic properties of metal-oxides are known to deviate from experimental results when evaluated with the generalized gradient approximation (GGA) to DFT.\cite{wang_oxidation_2006} Wang \textit{et al.} have identified the overbinding of the O$_{2}$ molecule and the self-interaction of localized \textit{d}-electrons in the transition metal as the main source of errors.\cite{wang_oxidation_2006} Various methods have been proposed to account for these errors.\cite{wang_oxidation_2006,wang_first-principles_2007,jain_formation_2011,grindy_approaching_2013} One straightforward approach to correct the O$_{2}$ overbinding is to add an empirical correction.\cite{wang_oxidation_2006,jain_high-throughput_2011,grindy_approaching_2013} We have adopted this approach for all GGA-based calculations by adding a recently proposed\cite{grindy_approaching_2013} correction of +1.20 eV to the total energy of O$_2$. Our calculated dissociation energy of O$_{2}$ is -4.85 eV after correction while the GGA uncorrected value is -6.05 eV; the corresponding experimental value is -5.13 eV.\cite{haynes_crc_2013} The self-interaction error can be treated with an onsite Hubbard model correction.\cite{anisimov_first-principles_1997} We used the rotational invariant approach of Dudarev,\cite{dudarev_electron-energy-loss_1998} where a Coulomb parameter U and exchange parameter J are combined into a single U-J parameter. For simplicity, we hereafter refer to U instead of U-J. The electronic states and energetics calculated with this method can depend on the chosen value of U.\cite{zhou_first-principles_2004,wang_oxidation_2006} To study how the properties of Mg-doped LiCoO$_{2}$ changes with U, we performed calculations with U = 1.5, 3.3, 5.0, and 5.5 eV. We used the U parameter consistently for Co in all calculations, including metallic Co. Note that within this DFT+U and empirical correction scheme the energies of s and p orbitals are not directly affected, leaving the well known GGA binding energy errors\cite{chevrier_hybrid_2010,stevanovic_correcting_2012,grindy_approaching_2013} untreated except for the case of the O$_2$ molecule.

\subsection{Chemical Potentials}

To evaluate Eq.~(\ref{eq:six}), we need to determine the atomic chemical potentials ${\mu}$$_{\textit{i}}$. The quaternary Li-Co-Mg-O phase diagram is available in the Materials Project\cite{jain_commentary:_2013,Ong2008,Jain2011a} website.\cite{comment3} From this phase diagram, one can expect separated Li-Co-O and MgO phases. Note that these are the known Li-Co-Mg-O phases but additional phases could exist. Based on this phase diagram, we approximate the atomic chemical potentials ${\mu}$$_{\textit{i}}$ assuming that LiCoO$_2$ is stable and in contact with MgO. 

The stability condition of LiCoO$_2$ requires that:

\begin{eqnarray}
\Delta\mu_{Li} + \Delta\mu_{Co} + 2\Delta\mu_{O}=\Delta H^f \left(LiCoO_2\right)
\label{eq:eight}
\end{eqnarray}

\noindent where $\Delta H^f$ is the formation enthalpy and $\Delta\mu_{Li}=\mu_{Li}^{LiCoO_2}-\mu_{Li}^{Li_{bulk}}$, $\Delta\mu_{Co}=\mu_{Co}^{LiCoO_2}-\mu_{Co}^{Co_{bulk}}$ and $\Delta\mu_{O}=\mu_{O}^{LiCoO_2}-\mu_{O}^{O_{2}(gas)}$. $\Delta\mu_{i}$ indicates\cite{wei_overcoming_2004,kramer_tailoring_2009,dixit_first-principles_2013} the possible variation of the chemical potential of atom \textit{i} when LiCoO$_2$ is formed, where $\mu_{i}^{LiCoO_2}$ characterizes the chemical potential of atom \textit{i} in LiCoO$_2$, and $\mu_{i}^{i_{bulk}}$ and $\mu_{O}^{O_{2}(gas)}$ the potential of atom \textit{i} in bulk and an O atom in O$_2$. Absence of segregation of bulk Li, Co or formation of O$_2$ gas implies that $\Delta\mu_{i}\le 0$. Additionally, LiCoO$_2$ competes with other possible Li-Co-O compounds, such as Li$_2$O, Li$_2$O$_2$, Li$_8$CoO$_6$, Li$_6$CoO$_4$, CoO and Co$_3$O$_4$.\cite{wang_first-principles_2007,kramer_tailoring_2009,koyama_defect_2012} Therefore, the chemical potentials are also constrained by: 

\begin{eqnarray}
x\Delta\mu_{Li} + y\Delta\mu_{Co} + z\Delta\mu_{O}\le\Delta H^f \left(Li_xCo_yO_z\right)
\label{eq:nine}
\end{eqnarray}

\noindent For MgO, the stability condition requires:

\begin{eqnarray}
\Delta\mu_{Mg} + \Delta\mu_{O}=\Delta H^f \left(MgO\right)
\label{eq:ten}
\end{eqnarray}

\noindent where $\Delta\mu_{Mg}=\mu_{Mg}^{MgO}-\mu_{Mg}^{Mg_{bulk}}$. 

After accounting for all constraints, the range of Li and O chemical potentials that stabilize LiCoO$_2$ are defined in the ($\Delta\mu_{Li}$, $\Delta\mu_{O}$) plane. For a given point in this plane, $\mu_{Co}^{LiCoO_2}$ is determined from Eq.~(\ref{eq:eight}) and $\mu_{Mg}^{MgO}$ from Eq.~(\ref{eq:ten}).

\section{Results and Discussion}

\subsection{Range of Chemical Potentials}

\begin{table*}
\caption{\label{tab:table1}Experimental\cite{wang_limo2_2005,chase_nist-janaf_1998} (at T = 298 K) and calculated formation enthalpy  of Li-Co-O compounds and MgO in eV. Results from GGA (U = 0) and GGA+U calculations with U = 1.5, 2.4, 3.3, 5.0 and 5.5 eV are included. Previous GGA and GGA+U results are included for comparison. The enthalpies of MgO, Li$_2$O, and Li$_2$O$_2$ are U independent.}
\begin{ruledtabular}
\begin{tabular}{llllllll}
System & Exp. & U=0 &  U=1.5 & U=2.4 & U=3.3 & U=5.0  & U=5.5 \\
 \hline
LiCoO$_2$   & -7.04   & -7.10  & -7.42 &  -7.61  & -7.79, -7.12,\footnote{Ref. \onlinecite{kramer_tailoring_2009}}  & -7.77, -6.97\footnote{Ref. \onlinecite{koyama_defect_2012}} & -7.73   \\
 &  &  & & & -7.13\footnote{Ref. \onlinecite{jain_commentary:_2013,Jain2011a,comment3}}  & &  \\
CoO   & -2.46   & -1.95, -1.9\footnote{Ref. \onlinecite{wang_oxidation_2006}}& -2.44 & -2.86 & -3.31, -2.47,$^\text{a}$  & -3.81,  -3.8$^\text{b}$& -3.94      \\
&  & & & & -2.66$^\text{c}$  & &  \\
Co$_3$O$_4$     & -9.43    & -9.36, -9.4$^\text{d}$ & -10.48 & -11.23 & -12.00, -9.76,$^\text{a}$ & -12.41,  -11.5$^\text{b}$ & -12.45     \\
&  & & & & -9.92$^\text{c}$ & & \\
Li$_6$CoO$_4$  &   & -20.66  & -21.38  & -21.82 & -22.26, -21.54$^\text{c}$  & -22.74, -20.5$^\text{b}$   & -22.87      \\
Li$_8$CoO$_6$  &   & -29.39  & -29.74  & -29.94 & -30.13, -29.50$^\text{c}$ & -30.14 &  -30.12      \\
Li$_2$O$_2$   & -6.56   & -6.97, -7.04$^\text{a}$ & &   &      \\
 &  & -6.60$^\text{c}$ & &   &      \\
Li$_2$O   & -6.21   & -6.21, -6.2,$^\text{d}$ &  & &  &      \\
&  & -6.28,$^\text{a}$ -5.5$^\text{b}$ &  &  & &      \\
&  & -6.21,$^\text{c}$ &  &  & &      \\
MgO   & -6.23   & -6.04, -6.1$^\text{d}$ & & &  & &      \\
  & & -6.14$^\text{c}$ & & &  & &      \\

\end{tabular}
\end{ruledtabular}
\end{table*}

To determine the chemical potential range of Li and O that stabilize LiCoO$_2$, we evaluated $\Delta H^f$ of stable Li-Co-O compounds.\cite{wang_first-principles_2007,kramer_tailoring_2009,koyama_defect_2012} The results from GGA and GGA+U are given in Table~\ref{tab:table1}. The formation enthalpy of MgO and available experimental and previously calculated $\Delta H^f$ values are also included in Table~\ref{tab:table1}.  Our calculated formation enthalpies are in general agreement with previous GGA and GGA+U calculations,\cite{wang_oxidation_2006,kramer_tailoring_2009,koyama_defect_2012} particularly for the non-transition metal-oxides and for GGA calculations. For the GGA+U results of Co-oxides some differences can be noticed. For instance, our calculated formation energy for LiCoO$_2$ with U = 3.3 and 5.0 eV is 0.7 - 0.8 eV lower than the values in Refs. \onlinecite{kramer_tailoring_2009, jain_high-throughput_2011,koyama_defect_2012}. The formation energy of CoO and Co$_3$O$_4$ evaluated with GGA+U also differs from Refs. \onlinecite{kramer_tailoring_2009, jain_high-throughput_2011} but resemble the results in Ref. \onlinecite{koyama_defect_2012}. The source of the discrepancy between our results and Ref. \onlinecite{kramer_tailoring_2009} is unclear. The discrepancy with the results of Ref. \onlinecite{koyama_defect_2012} arises mainly because a correction factor for the O$_{2}$ overbinding was not used in that work.

\begin{figure}
\includegraphics[scale=0.9]{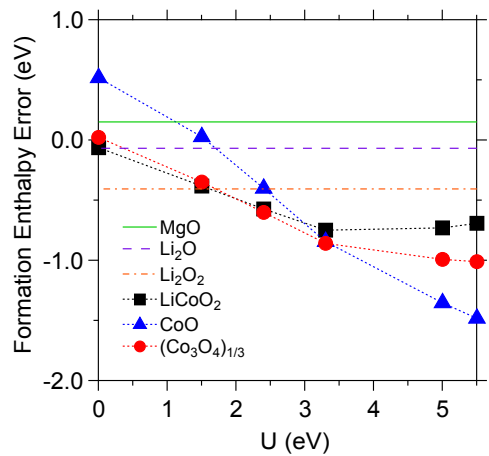}
\caption{\label{fig:fig2} Deviation between experimental and calculated formation enthalpies as function of U. Note that for the Co-oxides the error is per Co atom.}%
\end{figure}

We discuss first the deviation of our calculated formation enthalpies from the available experimental results to have an overall idea on the accuracy of the calculated chemical potentials. Note that the calculated formation enthalpies are at T= 0 K while the experimental results are at T = 298 K. However, the difference between 0 K and 298 K enthalpies is relatively small, within 0.1 eV per mol of O$_2$ for most transition metal-oxides.\cite{wang_oxidation_2006} Fig.~\ref{fig:fig2} shows the deviation between calculated and experimental formation enthalpy. GGA reproduces the formation enthalpy of non-transition metal-oxides (Li$_2$O, Li$_2$O$_2$, and MgO) after correcting for the O$_2$ overbinding.\cite{wang_oxidation_2006,grindy_approaching_2013} The error on the formation enthalpy of Li$_2$O and MgO is within 0.2 eV. For Li$_2$O$_2$, the error is 0.41 eV.

The formation enthalpy of Co-oxides evaluated within GGA are expected to deviate from experiment due to the self-interaction of localized \textit{d}-electrons.\cite{zhou_first-principles_2004, wang_oxidation_2006} The deviation on the formation enthalpy of CoO evaluated with GGA is close to 0.50 eV. For LiCoO$_2$ and Co$_3$O$_4$, the corresponding deviation is relatively small, less than 0.1 eV. However, the oxidation energy of CoO (6CoO + O$_2$ $\rightarrow$ 2Co$_3$O$_4$) is -7.04 eV, when evaluated with GGA, while the experimental value is -4.08 eV.  GGA+U with U = 1.5 eV leads to a lower deviation of the formation enthalpy of CoO (0.03 eV) from experiment, but a greater deviation for LiCoO$_2$ (0.38 eV) and Co$_3$O$_4$ (1.19 eV). The oxidation energy of CoO is also large, -6.34 eV, a 2.26 eV error. The agreement with experiment for the oxidation of CoO is better when it is calculated with U = 3.3 eV;\cite{wang_oxidation_2006} the calculated oxidation energy is -4.15 eV, only an 0.07eV error. However, the error in the formation enthalpy of LiCoO$_2$, CoO and Co$_3$O$_4$ is above 0.70 eV when evaluated with U = 3.3 eV. As shown in Fig.~\ref{fig:fig2}, the agreement for the oxidation energy of CoO may be due to error cancelation because the error per Co atom of the formation energy of CoO and Co$_3$O$_4$ is the same for U = 3.3 eV. For U values over 3.3 eV, the formation enthalpies and relative energies of CoO and Co$_3$O$_4$ deviate from experimental results by more than 1 eV.

The enthalpy of formation of LiCoO$_2$ is less sensitive to the U value than CoO  and Co$_3$O$_4$. For U values from 3.3 to 5.5 eV, the deviation on the formation enthalpy of LiCoO$_2$ is centered on 0.7 eV. U values in this range are commonly used to study different processes and properties of LiCoO$_2$. U = 2.9 eV was used to study the electronic structure of LiCoO$_2$.\cite{ensling_electronic_2010} A U value of 3.3 eV has been used to study the phase diagram and surface properties of LiCoO$_2$.\cite{kramer_tailoring_2009} This is also the value adopted for Co in the Materials Project.\cite{jain_high-throughput_2011} This value of U was established from a fit to the experimental oxidation energy of CoO and the methodology of Wang \textit{et al.}\cite{wang_oxidation_2006} to correct for the O$_2$ overbinding. U values close to 5.0 eV were used to calculate the average Li-intercalation potential,\cite{zhou_first-principles_2004,chevrier_hybrid_2010} and the defect chemistry\cite{koyama_defect_2012} in LiCoO$_2$. U = 5.5 eV has been used to study the phase diagram and thermal decomposition of LiCoO$_2$.\cite{wang_first-principles_2007} U values close to 5.0 eV or 5.5 eV are taken from the self-consistently determined\cite{zhou_first-principles_2004}  U value of Co in layered LiCoO$_2$ or the average of U values of Co in different  Co-oxides. 

Clearly, the GGA+U method is limited in accuracy for relative energies of metal-oxides, even when empirically choosing U. This is not unexpected as the U parameter should be sensitive to the chemical environment of the atom sites where the correction is applied.\cite{zhou_first-principles_2004} To calculate formation enthalpy of metal-oxides, the main problem is using the same U value to describe the atom sites in the metallic state (reactant) and the oxides (product). The method proposed by Jain \textit{et al.}\cite{jain_formation_2011} and used in the Material Project is a possible solution to this problem. Formation enthalpies evaluated with this method are in reasonable agreement with experiments.\cite{jain_formation_2011} Yet, Jain's method relies on U values determined to describe formation enthalpies or oxidation energy of metal-oxides. This limits the usefulness of the method to study defects, as the U values are not necessarily transferable. As exemplified by LiCoO$_2$, a single U value cannot describe all properties, not even formation enthalpies and oxidation energies of the known parent phases of transition metal oxides to reasonable accuracy ($< 0.5$ eV).

\begin{figure}
\includegraphics[scale=0.9]{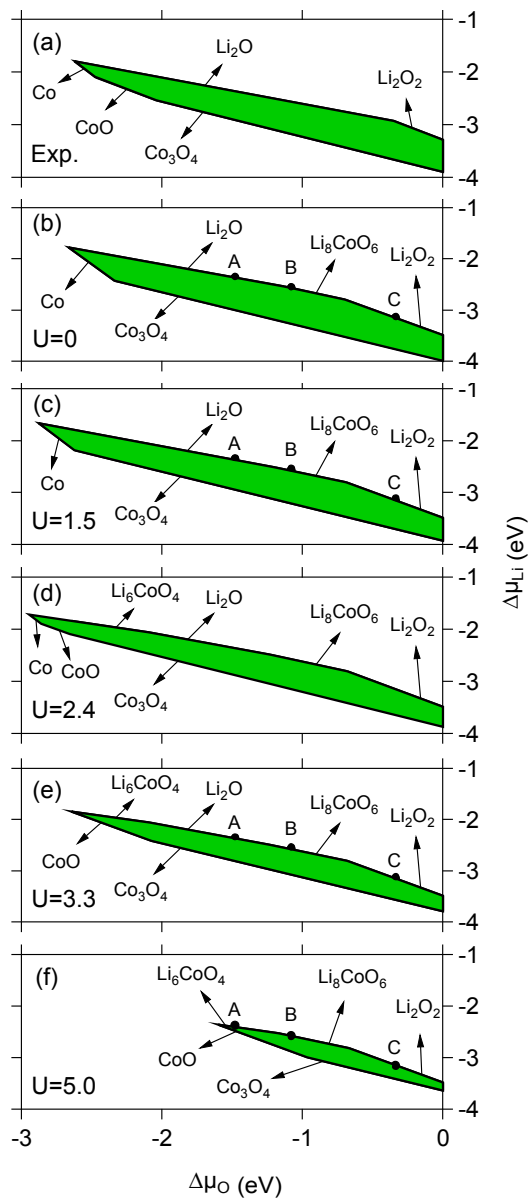}
\caption{\label{fig:fig3} Boundaries of the stable region of LiCoO$_2$ in the O and Li chemical potentials. The boundaries were constructed with the formation enthalpy from (a) experiment, (b) GGA and GGA+U with U values of (c) 1.5, (d) 2.4, (e) 3.3 and (f) 5.0 eV. The filled circles A, B, and C indicates the chemical potentials where defect formation energies were evaluated.}%
\end{figure}

We now discuss the phase stability diagram of LiCoO$_2$ and competing phases evaluated with GGA and GGA+U (Fig.~\ref{fig:fig3}). We have limited our analysis to the listed Li-Co-O compounds as these are the known or expected\cite{wang_first-principles_2007} thermally stable phases. For each diagram in Fig.~\ref{fig:fig3}, the colored polygon shows the range of O and Li chemical potentials that stabilize LiCoO$_2$. The reference for the chemical potential of O and Li is the O$_2$ molecule at 0 K and metallic lithium, respectively. The chemical potential of gaseous oxygen is dependent on temperature and partial pressure. One can approximate the chemical potential of gaseous oxygen assuming it to be an ideal gas on the basis of experimental data.\cite{haynes_crc_2013} For instance, at ambient pressure and 1200 K, 900 K and 298 K, it is approximately 1.5, 1.0 and 0.3 eV below $\mu_{O}^{O_{2}(gas)}$.

Fig.~\ref{fig:fig3}(a) shows the chemical potential diagram constructed from available experimental formation enthalpy of Li-Co-O compounds at  T = 298 K.\cite{wang_limo2_2005,chase_nist-janaf_1998} Fig.~\ref{fig:fig3}(b), (c), (d), (e) and (f) correspond to the phase diagrams from GGA and GGA+U calculations with U = 1.5, 2.4, 3.3 and 5.0 eV, respectively. All calculations predict a range of O and Li chemical potentials where LiCoO$_2$ is stable. Yet, the potential ranges differ from experiment and are noticeably sensitive to the U value. For instance, the maximum oxygen chemical potential where LiCoO$_2$ is stable change from -2.88 eV for calculations with U = 1.5 eV to only -1.60 eV for U = 5.0 eV. Moreover, only calculations with U $\sim 2.4$ eV reproduces all the phases that are in equilibrium with LiCoO$_2$. GGA and GGA+U with U = 1.5 eV predict the equilibrium between metallic Co and LiCoO$_2$ phases but not the equilibrium with the CoO phase (i.e. it is missing from the phase diagram). On the other hand, GGA+U with a U value of 3.3 eV or higher show the equilibrium of LiCoO$_2$ with CoO but not with metallic Co.

Based on these phase diagrams, we studied the extrinsic Mg defects in LiCoO$_2$ under three different chemical conditions, labeled as A, B and C. The chemical conditions A, B, and C corresponds to Li-rich condition at 0.2 atm oxygen and 1200 K, 900 K and 298 K, respectively. These conditions are indicated as filled circles in Fig.~\ref{fig:fig3}.

\subsection{Defect Formation Energies}

\begin{figure}
\includegraphics[scale=0.9]{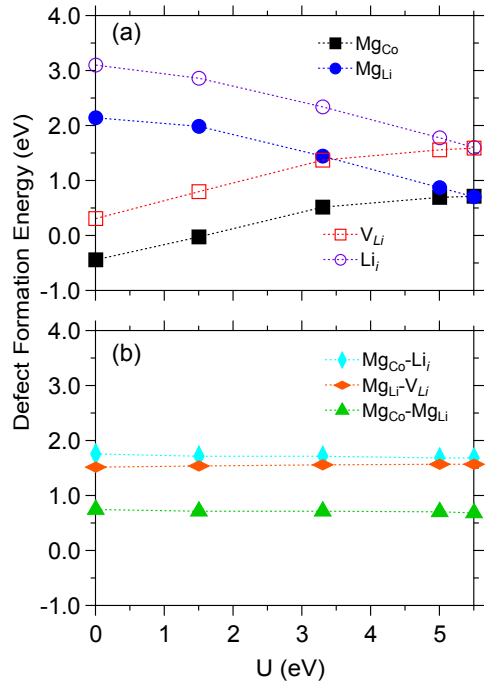}
\caption{\label{fig:fig4} Formation energy of (a) Mg and Li defects and (b) their complexes in LiCoO$_2$ as function of U. The energies are obtained at point A in the chemical potential diagram. Note overlapping symbols in  (a) at the value U = 5.5.}%
\end{figure}

The electronic species formed upon the substitution of Mg on Co and Li sites can be in one of multiple spin configurations. To identify the configuration with the lowest energy, we examined various spin configurations for each defect. Similarly, spin configurations were evaluated for the Li$_{\textit{i}}$ and V$_{Li}$ defects. The defect complexes of Mg have non-polarized spin configurations. The formation energy of Mg$_{Co}$, Mg$_{Li}$, Mg$_{Co}$-Li$_{\textit{i}}$, Mg$_{Li}$-V$_{Li}$ and Mg$_{Co}$-Mg$_{Li}$ are shown in Fig.~\ref{fig:fig4} as a function of U for a representative chemical condition. We also included the formation energy of Li$_{\textit{i}}$ and V$_{Li}$. The chemical condition correspond to point A in Fig.~\ref{fig:fig3}, representing the most Li-rich condition during synthesis at 1200 K.

As previously discussed, the formation of Mg$_{Co}$ and Mg$_{Li}$ leads to changes on the valency of Co atoms in LiCoO$_2$. As a result, the formation energy of these defects depends on the U value, Fig.~\ref{fig:fig4}(a). The change in the slope of the formation energy of Mg$_{Co}$ and Mg$_{Li}$ at U = 3.3 and 1.5 eV, respectively, corresponds to a transition from delocalized (low U) to localized (high U) states; see discussion in the next section. The same U dependence is also found in the formation energy of the V$_{Li}$ and Li$_{\textit{i}}$ point defects since the valency of Co atoms also changes upon formation of these defects. The result for the formation energy of V$_{Li}$ as function of U resembles previous calculations of the Li-insertion voltage in LiCoO$_2$.\cite{zhou_first-principles_2004} In contrast to Mg$_{Co}$ and Mg$_{Li}$ defects, formation of the Mg defect complexes do not change the valency of Co in LiCoO$_2$ and their formation energies are rather insensitive to U, Fig.~\ref{fig:fig4}(b).

These results show the difficulty of studying Mg doping in LiCoO$_2$ employing GGA and GGA+U calculations. For instance, the preferred site location of Mg changes with the U value. Mg is preferentially located on Co sites for calculations with GGA and GGA+U with U below 5.0 eV, which is in line with experiments\cite{tukamoto_electronic_1997,julien_combustion_2000,levasseur_dual_2002,thirunakaran_mg_2003,kim_electrochemical_2004,kim_crystal_2006,sathiyamoorthi_layered_2007,eom_m3_2008,zaheena_microwave_2009,zhou_relative_2009,luo_synthesis_2010,valanarasu_effect_2010} showing that Mg is located at Co site. For calculations with U = 5.0 and 5.5 eV, however, Mg is equally stable on both Co and Li sites. Moreover, only GGA+U with U values around 1.5 eV predict the spontaneous formation of Mg-doped LiCoO$_2$. The formation energy of Mg$_{Co}$ is -0.44 eV and -0.02 eV for GGA and GG+U with U = 1.5 eV, respectively. GGA+U calculations with U values of 3.3, 5.0 and 5.5 eV yield Mg$_{Co}$ formation energies of 0.52, 0.70 and 0.72 eV, respectively, predicting that Mg will have a low solubility in LiCoO$_2$. However, Mg-doped LiCoO$_2$ with Mg/Co ratio as high as 0.5 have been synthesized,\cite{sathiyamoorthi_layered_2007} demonstrating that this is not the case.

\begin{table*}[!ht]
\tiny
\caption{\label{tab:table2}Calculated formation energy of Mg and Li defects and binding energy (BE) of their complexes in LiCoO$_2$ and in eV. Results from GGA (U = 0) and GGA+U calculations with U = 1.5, 3.3, 5.0 and 5.5 eV are included. The points A, B and C indicates the chemical potentials that were examined, corresponding to Li-rich condition and 0.2 atm O$_2$ at A: 1200 K, B: 900 K and C: 298 K.} 
\begin{ruledtabular}
\begin{tabular}{lcccccccccccccccccccc}
&\multicolumn{4}{c}{U=0}  &\multicolumn{4}{c}{U=1.5} &\multicolumn{4}{c}{U=3.3}  &\multicolumn{4}{c}{U=5}  &\multicolumn{4}{c}{U=5.5}  \\

Defects  & A & B & C & BE & A & B & C & BE & A & B & C & BE  & A & B & C & BE & A & B & C & BE \\
\hline
Mg$_{Co}$ &	-0.44 &	-0.64 &	-0.80 &	 &	-0.02 &	-0.22 &	-0.38 &	 &	0.52 &	0.32 &	0.16 &	 &	0.70 &	0.50 &	0.34 &	 &	0.72 &	0.52 &	0.36 &		\\
Mg$_{Li}$  &	2.15 &	2.34 &	2.50 &	 &	1.99 &	2.19 &	2.35 &	 &	1.45 &	1.66 &	1.82 &	 &	0.88 &	1.08 &	1.24 &	 &	0.71 &	0.91 &	1.07 &		\\
Li$_{\textit{i}}$ &	3.10 &	3.30 &	3.88 &	 &	2.86 &	3.06 &	3.64 &	 &	2.34 &	2.54 &	3.12 &	 &	1.78 &	1.98 &	2.57 &	 &	1.60 &	1.80 &	2.38 &		\\
V$_{Li}$  &	0.31 &	0.11 &	-0.48 &	 &	0.80 &	0.60 &	0.02 &	 &	1.38 &	1.18 &	0.60 &	 &	1.56 &	1.36 &	0.78 &	 &	1.59 &	1.39 &	0.81 &		\\
Mg$_{Co}$-Li$_{\textit{i}}$ &	1.76 &	1.76 &	2.18 &	0.90 &	1.72 &	1.72 &	2.14 &	1.12 &	1.72 &	1.72 &	2.14 &	1.14 &	1.69 &	1.69 &	2.11 &	0.79 &	1.69 &	1.69 &	2.11 &	0.63	\\
Mg$_{Li}$-V$_{Li}$ &	1.52 &	1.52 &	1.10 &	0.92 &	1.54 &	1.54 &	1.12 &	1.24 &	1.56 &	1.56 &	1.14 &	1.28 &	1.57 &	1.57 &	1.14 &	0.88 &	1.57 &	1.57 &	1.14 &	0.73	\\
Mg$_{Co}$-Mg$_{Li}$ &	0.75 &	0.75 &	0.75 &	0.95 &	0.72 &	0.72 &	0.72 &	1.24 &	0.72 &	0.72 &	0.72 &	1.25 &	0.71 &	0.71 &	0.71 &	0.87 &	0.69 &	0.69 &	0.69 &	0.74	\\
\end{tabular}
\end{ruledtabular}
\end{table*}

The formation energy of the Mg defects in LiCoO$_2$ depend on the atomic chemical potentials. We list in Table~\ref{tab:table2} the formation energy of Mg defects in LiCoO$_2$ for various chemical potentials. These potentials represent Li-rich environment and 0.2 atm O$_2$ at (A) 1200, (B) 900 and (C) 298 K. The formation energy of Mg$_{Co}$ is reduced from point A to B and C while that of Mg$_{Li}$ increases. Only GGA+U with U = 1.5 eV predicts Mg to be easily soluble in LiCoO$_2$ at points A, B and C. For greater U values, Mg$_{Co}$ is energetically unfavorable at all points.

Table ~\ref{tab:table2} also includes the formation and binding energy of the Mg$_{Co}$-Li$_{\textit{i}}$, Mg$_{Li}$-V$_{Li}$ and Mg$_{Co}$-Mg$_{Li}$  complexes. The Mg$_{Co}$-Li$_{\textit{i}}$ and Mg$_{Li}$-V$_{Li}$ complexes have high formation energies, from 1.1 to 2.2 eV at the various atomic chemical potentials and binding energy from 1.3 to 0.6 eV. The Mg$_{Co}$-Mg$_{Li}$ complex has a formation energy close to 0.7 eV that is independent of the chemical potential and with binding energies from 0.7 to 1.3 eV. Although the formation energy of Mg$_{Co}$-Mg$_{Li}$ is high, it is the lowest for U=5.5 eV at point A and often within a few 0.1 eV elsewhere. The formation energy of these complexes is insensitive to the U value and we can state with some certainty that these complexes are unlikely to form in Mg-doped LiCoO$_2$ due to their high formation energies.

\subsection{Defect Geometries}

To study the local geometry around Mg in LiCoO$_2$ as a function of U, we have tabulated in Table~\ref{tab:table3} the Mg-O and Co-O interatomic distances calculated with GGA and GGA+U. For the Mg$_{Co}$ and Mg$_{Li}$ defects, we also include the magnetic moment of the supercell and whether it is localized or delocalized. The localization was determined based on the local magnetic moments. 

For Mg$_{Co}$, both GGA and GGA+U predict a low-spin configuration. Yet, for calculations with U = 3.3 or below, the state is delocalized while for U =  5.0 eV or above, it is localized. When the state is delocalized, the Mg-O distances are 2.04 $\mathring{\textnormal{A}}$ independently of the U value (Table~\ref{tab:table3}). The Co-O distances of Co atoms nearest to Mg$_{Co}$ are 1.94 $\mathring{\textnormal{A}}$, as in pristine LiCoO$_2$, when calculated with GGA. For GGA+U with U = 1.5 and 3.3 eV, these Co-O bonds are slightly distorted with distances of 1.94 and 1.93 $\mathring{\textnormal{A}}$. GGA+U results in a state that is delocalized mainly on the six Co atoms nearest to Mg$_{Co}$ while GGA leads to a more delocalized state and undistorted Co-O bonds.

For the localized state predicted with U = 5.0 eV, the Mg-O bonds are distorted with distances ranging from 2.06 to 2.02 $\mathring{\textnormal{A}}$. A slightly higher U value, 5.5 eV, leads to higher distortion with Mg-O distances ranging from 2.07 to 1.95 $\mathring{\textnormal{A}}$. This distortion on the Mg-O bonds results from the formation of the localized holes on a Co atom near to Mg$_{Co}$. The Co-O bonds where the hole is located are also distorted with distances from 1.92 to 1.90 $\mathring{\textnormal{A}}$ (Table~\ref{tab:table3}). Such bonds are shorter than 1.94 $\mathring{\textnormal{A}}$ for Co-O in pristine LiCoO$_2$. 

\begin{table*}
\tiny
\caption{\label{tab:table3}Total magnetic moment (\textit{MM} per supercell in bohr magneton $\mu_{B}$) and Mg-O and Co-O interatomic distances ($\mathring{\textnormal{A}}$) in Mg-doped LiCoO$_2$.  Results are included for GGA and GGA+U calculations with U = 1.5, 3.3, 5.0 and 5.5 eV.}
\begin{ruledtabular}
\begin{tabular}{llllllllllllll}
Defects & Property & U=0 & U=1.5 & U=3.3 & U=5 & U=5.5  \\
\hline
Mg$_{Co}$  & \textit{MM} & 1 - \textit{delocalized}  & 1 - \textit{delocalized}  & 1 - \textit{delocalized} & 1 - \textit{localized} & 1 - \textit{localized}  \\
 & \textit{d}(Mg-O) & 2.04$\times$6 & 2.04$\times$6 & 2.04$\times$6 & 2.06$\times$2, 2.03$\times$2, & 2.07, 2.04, 2.03$\times$2,  \\
 & & & & & 2.02$\times$2 &  2.01, 1.95 \\
 &  \textit{d}(Co-O) & 1.94$\times$6 & 1.94$\times$4, 1.93$\times$2 & 1.94$\times$4, 1.93$\times$2 & 1.92$\times$2, 1.91$\times$2, & 1.92$\times$2, 1.91$\times$2,  \\
 & & & & & 1.90$\times$2 & 1.87$\times$2  \\
V$_{Li}$ & \textit{MM} & 1 - \textit{delocalized} & 1 - \textit{delocalized} & 1 - \textit{delocalized} & 1 - \textit{localized} & 1 - \textit{localized}  \\
 &  \textit{d}(Co-O) & 1.95$\times$2, 1.94$\times$2, & 1.94$\times$2, 1.93$\times$2, & 1.94$\times$2, 1.93$\times$2,  & 1.92$\times$2, 1.91$\times$2, & 1.92$\times$2, 1.91$\times$2,  \\
 &  & 1.92$\times$2 & 1.92$\times$2 & 1.92$\times$2  & 1.89$\times$2 & 1.89$\times$2  \\
Mg$_{Li}$ & \textit{MM} & 1 - \textit{delocalized} & 3 - \textit{localized} & 3 - \textit{localized} & 3 - \textit{localized} & 3 - \textit{localized}  \\
 &  \textit{d}(Mg-O) & 2.08$\times$6 & 2.09$\times$2, 2.08$\times$2 & 2.10$\times$2, 2.09, & 2.10$\times$2, 2.09, & 2.10$\times$2, 2.09,   \\
 & & & 2.07, 2.01  & 2.08, 2.06$\times$2 & 2.08, 2.06$\times$2  & 2.08, 2.06$\times$2  \\
 &  \textit{d}(Co-O) & 1.96$\times$2, 1.94$\times$2, & 2.08, 2.06$\times$2, & 2.10$\times$2, 2.06, & 2.10$\times$2, 2.06,  & 2.10$\times$2, 2.06,   \\
 &  & 1.93$\times$2 & 2.05$\times$3 & 2.05, 2.03$\times$2 & 2.05, 2.04$\times$2 & 2.05, 2.04$\times$2  \\
Li$_{\textit{i}}$ & \textit{MM} & 1 - \textit{delocalized} & 3 - \textit{localized} & 3 - \textit{localized} & 3 - \textit{localized} & 3 - \textit{localized}  \\
 &  \textit{d}(Co-O) & 1.98$\times$2, 1.97, & 2.18$\times$2, 2.14,  & 2.16$\times$2, 2.13,  & 2.14$\times$3, 2.00$\times$2, & 2.15, 2.14, 2.13, \\
 & & 1.92$\times$2, 1.91 & 1.98, 1.97$\times$2 & 2.00, 1.99, 1.98 & 1.98 &  2.00, 1.99$\times$2  \\
Mg$_{Co}$-Li$_{\textit{i}}$	  &   \textit{d}(Mg-O)	  &  2.11$\times$3, 1.98$\times$3	  &  2.11$\times$3, 1.98$\times$3	  &  2.11$\times$3, 1.98$\times$3	  &  2.11$\times$3, 1.98$\times$3	  &  2.11$\times$3, 1.98$\times$3	\\
	  &   \textit{d}(Co-O)	  &  1.95, 1.94$\times$2, 	  &  1.94$\times$3, 1.93$\times$2,	  &  1.94$\times$3, 1.93$\times$2,	  &  1.94$\times$3, 1.93$\times$2,	  &  1.94$\times$3, 1.93$\times$2,	\\
	  &  	  &  1.93$\times$2, 1.92	  &  1.92	  &  1.92	  &  1.92	  &  1.92	\\
Mg$_{Li}$-V$_{Li}$	  &   \textit{d}(Mg-O)	  &  2.12$\times$2, 2.08$\times$2,	  &  2.12$\times$2, 2.08$\times$2,	  &  2.12$\times$2, 2.08$\times$2, 	  &  2.12$\times$2, 2.08$\times$2, 	  &  2.12$\times$2, 2.08$\times$2, 	\\
	  &  	  &  2.06$\times$2 	  &  2.06$\times$2	  &  2.05$\times$2	  &  2.05$\times$2	  &  2.05$\times$2	\\
	  &   \textit{d}(Co-O)	  &  1.95$\times$2, 1.94$\times$2, 	  &  1.95$\times$2, 1.94$\times$2,	  &  1.95$\times$2, 1.94$\times$2,	  &  1.95$\times$2, 1.94$\times$2, 	  &  1.95$\times$2, 1.94$\times$2, 	\\
	  &  	  &  1.92$\times$2	  &  1.92$\times$2	  &  1.92$\times$2	  &   1.92$\times$2, 	  &  1.92$\times$2	\\
Mg$_{Co}$-Mg$_{Li}$	  &   \textit{d}(Mg$_{Co}$-O)	  &  2.07$\times$2, 2.04$\times$2,	  &  2.07$\times$2, 2.04$\times$2,	  &  2.07$\times$2, 2.04$\times$2,	  &  2.07$\times$2, 2.04$\times$2,	  &  2.07$\times$2, 2.04$\times$2,	\\
	  &  	  &  2.02$\times$2	  &  2.02$\times$2	  &  2.02$\times$2	  &  2.02$\times$2	  &  2.01$\times$2	\\
	  &   \textit{d}(Mg$_{Li}$ -O)	  &  2.10$\times$2, 2.09, 	  &  2.10$\times$2, 2.09, 	  &  2.10$\times$2, 2.09, 	  &  2.10$\times$2, 2.09, 	  &  2.10$\times$2, 2.09, 	\\
	  &  	  &  2.08, 2.06$\times$2 	  &  2.08, 2.06$\times$2 	  &  2.08, 2.06$\times$2 	  &  2.08, 2.06$\times$2	  &  2.08, 2.06$\times$2	\\
\end{tabular}
\end{ruledtabular}
\end{table*}

The impurity hole formed with Mg$_{Co}$ in Mg-doped LiCoO$_2$ can be compared with other hole that can be formed in LiCoO$_2$. We performed calculations for the hole created with the formation of a Li vacancy; results are included in Table~\ref{tab:table3}. As for Mg$_{Co}$, both GGA and GGA+U predict a low-spin configuration, delocalized for calculations with U = 3.3 (or below) and localized for U =  5.0 eV (or above). Another hole is the one formed as an intrinsic electronic defect in pristine LiCoO$_2$. The formation of this state in LiCoO$_2$ was recently studied\cite{koyama_defect_2012} with GGA+U (U = 5.0 eV) and methodologies similar to the present work. We therefore used this result for comparison. The intrinsic hole in LiCoO$_2$ was found to be localized with a low-spin configuration and Co-O distances of 1.91($\times$4) and 1.90($\times$2) $\mathring{\textnormal{A}}$.\cite{koyama_defect_2012} The properties of this hole are similar to the one formed with Mg$_{Co}$; the main difference is the more distorted Co-O geometry in Mg-doped LiCoO$_2$. 

For the impurity electron formed with the Mg$_{Li}$ defect, GGA predicts a delocalized low-spin state. The Mg-O distances are all 2.08 $\mathring{\textnormal{A}}$ while the Co-O bonds are distorted with interatomic distances from 1.96 to 1.93 $\mathring{\textnormal{A}}$. The Li-O interatomic distance in pristine LiCoO$_2$ is 2.11 $\mathring{\textnormal{A}}$. All calculations of the Mg$_{Li}$ defect with GGA+U result on a localized high-spin configuration. In this case, the Mg-O bonds are distorted with distances ranging from 2.10 to 2.01 $\mathring{\textnormal{A}}$. The impurity electron is localized on a second nearest neighbor Co atom to Mg$_{Li}$ with Co-O distances that range from 2.10 to 2.03 $\mathring{\textnormal{A}}$. These distances do not change significantly for calculations with the various U values.

The electron introduced with the Mg$_{Li}$ defect is comparable to the electron introduced upon the formation of an interstitial Li atom in LiCoO$_2$. GGA predicts a delocalized low-spin state and GGA+U a delocalized high-spin state for Li$_{\textit{i}}$ as for the Mg$_{Li}$ defect. The Co-O interatomic distances are also rather similar in both defects. For the electron as an intrinsic electronic defect in pristine LiCoO$_2$, Koyama  \textit{et al.}\cite{koyama_defect_2012} found a localized low-spin state with Co-O distances of 2.06 ($\times$6) $\mathring{\textnormal{A}}$.

In line with results for the formation energy, the characterization of the local geometry of defects that change the valency of Co atoms in LiCoO$_2$ is rather difficult with GGA+U calculations. The geometry of these defects is sensitive to the U value because increasing/decreasing U leads to localized/delocalized defect states (see discussion below).  The situation is different for defect complexes. The geometry of the Mg defect complexes in LiCoO$_2$ does not depend on the U value. For all complexes the Mg-O and Co-O bonds are distorted with distances from 2.11 to 1.93 $\mathring{\textnormal{A}}$ for Mg-O and from 1.95 to 1.92 $\mathring{\textnormal{A}}$ for Co-O (Table~\ref{tab:table3}). For these type of defects that do not change the valency of Co, GGA+U calculations yield more consistent geometric properties. 

\subsection{Electronic Structure}

\begin{figure}
\includegraphics[scale=0.9]{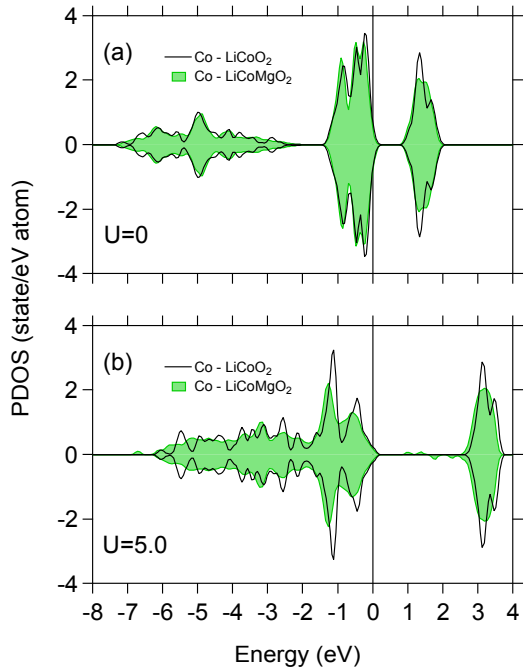}
\caption{\label{fig:fig5} Projected density of states (PDOS) of Co in LiCoO$_2$ (solid curve) and Mg-doped LiCoO$_2$ (green colored area). Panels (a) and (b) correspond to GGA and GGA+U (U = 5.0 eV) calculations of the Mg$_{Co}$ extrinsic defect. The energy is relative to the Valence Band Maximum (VBM).}%
\end{figure}

LiCoO$_2$ is a wide-band gap semiconductor. As found in previous calculations,\cite{czyzyk_band-theory_1992,xu_electronic_2003,shi_effect_2007,ensling_electronic_2010,juhin_angular_2010,sun_theoretical_2011,okumura_correlation_2012} GGA underestimates the band gap from the Valence Band Maximum (VBM) to the Conduction Band Minimum (CBM), i.e. 0.94 eV. The experimental values are 2.1,\cite{kushida_narrowing_2002} 2.5\cite{rosolen_photoelectrochemical_2001} and 2.7 eV.\cite{van_elp_electronic_1991} For GGA+U calculations, the predicted band gap depends on the U value.  The band gap calculated with U = 3.3 eV is 2.18 eV, in agreement with the experimental value of 2.1 eV in Ref. \onlinecite{kushida_narrowing_2002}, while the value of U = 5.0 eV, 2.76 eV, reproduces the experimental value of 2.7 eV in Ref. \onlinecite{van_elp_electronic_1991}. These GGA+U results are similar to previous calculations with U values close to 3\cite{ensling_electronic_2010} and 5 eV.\cite{ensling_electronic_2010,juhin_angular_2010,koyama_defect_2012,xiong_atomic_2012,hoang_defect_2014} 

The electronic structure of Mg-doped LiCoO$_2$ was studied by calculating the electronic band structure and density of states (DOS) with GGA and GGA+U. Fig.~\ref{fig:fig5} displays the projected DOS (PDOS) of Co in LiCoO$_2$ and Mg-doped LiCoO$_2$. The PDOS of Co in Fig.~\ref{fig:fig5}(a) and (b) can be ascribed to three main groups.\cite{aydinol_ab_1997} The peaks from -7 to -2 eV corresponds to the occupied valence bands e$_{g}$$^{b}$, the group from -2 to 0.5 eV to the partially occupied valence band t$_{2g}$ and the peaks from 0.5 to 4 eV to the unoccupied conduction band e$_{g}$$^{*}$.\cite{aydinol_ab_1997} 

The PDOS of Co in Mg-doped LiCoO$_2$ calculated with GGA, Fig.~\ref{fig:fig5}(a), is rather similar to that in pristine LiCoO$_2$. Yet, some differences can be noticed. After Mg-doping, the PDOS in the energy range of the e$_{g}$$^{b}$ bands slightly increase while the PDOS in the region of the t$_{2g}$ bands decrease. In the region of the e$_{g}$$^{*}$ bands, the PDOS also decrease. These changes in the electron distribution upon doping with Mg indicate that \textit{d}-electrons move from nonbonding bands to bonding bands.\cite{xu_electronic_2003} The result is a higher valence state for Co in Mg-doped LiCoO$_2$ than in LiCoO$_2$. In turn, the O anions become more closed-shell like as indicated by the decrease of the PDOS in e$_{g}$$^{*}$ bands region.\cite{xu_electronic_2003} The electron distribution also changes if Mg is located on a Li site instead of Co (data not shown). Moreover, these changes are also observed on the PDOS of Co calculated with GGA+U , Fig.~\ref{fig:fig5}(b). Therefore, both GGA and GGA+U calculations suggest that the valence of Co in LiCoO$_2$ is modified when doping with Mg.

\begin{figure}
\includegraphics[scale=0.9]{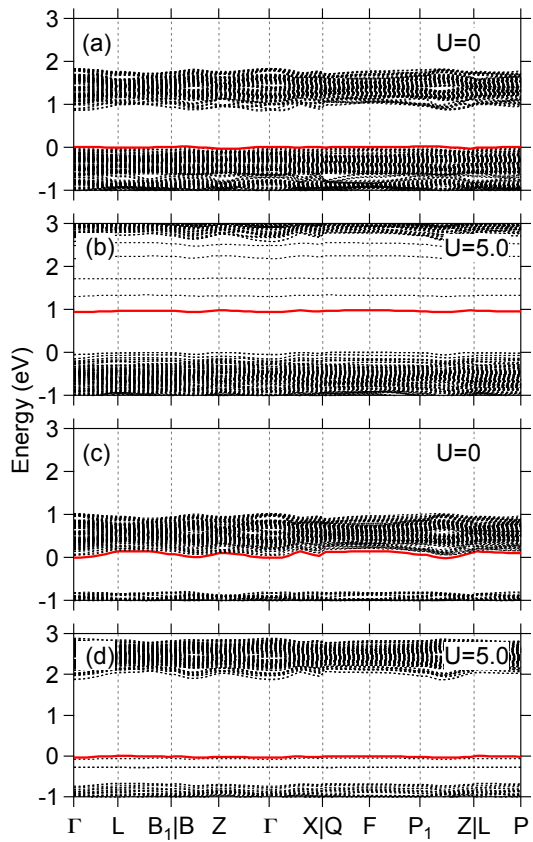}
\caption{\label{fig:fig6} Electronic band structures in the energy gap region of LiCoO$_2$ with the Mg$_{Co}$ (a and b) and Mg$_{Li}$ defect (c and d) calculated with (a and c) GGA and (b and d) GGA+U with U = 5.0. The solid red curve indicates the impurity band. The energy is relative to the Valence Band Maximum (VBM). The high-symmetry directions of the rhombohedral Brillouin zone were generated with AFLOW.\cite{setyawan_high-throughput_2010} }%
\end{figure}

As discussed previously, Mg$_{Co}$ in LiCoO$_2$ introduces an impurity hole in the system. This hole will be manifested in the band structure as an empty band close to the top of the valence band (shallow acceptor) or a band located "deep" in the band gap region. On the other hand, the formation of Mg$_{Li}$ results in a electron that will lead to an occupied state close to the bottom of the conduction band (shallow donor) or an impurity state in the band gap region. In principle, both the shallow acceptor and shallow donor can increase the carrier density, leading to higher electronic conductivity. Conversely, no changes in electronic conductivity are expected if the impurity state is located in the band gap. Since Mg-doped LiCoO$_2$ have been shown\cite{carewska_electrical_1997,tukamoto_electronic_1997}  to have up to 2 orders of magnitude higher electronic conductivity than LiCoO$_2$, one can expect either shallow acceptor or donor states in the band structure of Mg-doped LiCoO$_2$.

The electronic band structure near the valence and conduction bands of LiCoO$_2$ with Mg$_{Co}$ and Mg$_{Li}$ is shown in Fig.~\ref{fig:fig6}. As expected, an empty band near the top of the valence band is predicted by GGA when Mg is located on Co, Fig.~\ref{fig:fig6}(a). Characteristic of a shallow acceptor level,\cite{van_de_walle_first-principles_2004} this level exhibits a similar dispersion as the upper valence band. A similar band structure is expected for GGA+U calculations with U = 1.5 and 3.3 eV since the geometry and spin-configuration of Mg$_{Co}$ calculated with these methods are rather similar to those from GGA; this is confirmed by an increase in the density of states near the valence band maximum (data not shown). For GGA+U calculations with U = 5.0 eV or above, the impurity band associated with Mg$_{Co}$ is located in the band gap region, Fig.~\ref{fig:fig6}(b). This level is spatially localized, leading to the local deformation observed for Co-O and Mg-O interatomic distances (Table~\ref{tab:table3}) and the splitting of unoccupied d-bands. When Mg is located on Li, calculation with GGA shows an occupied band near the conduction band, Fig.~\ref{fig:fig6}(c). Notice that this band is now positioned at the Fermi level (or VBM). This shallow donor level exhibits dispersion similar to the bottom conduction band. As in Mg$_{Co}$, GGA+U leads to a localized impurity band when Mg$_{Li}$ is formed, Fig.~\ref{fig:fig6}(d).

The energy position of the impurity levels in Mg-doped LiCoO$_2$ are sensitive to the U value. For the Mg$_{Co}$ defect, GGA and GGA+U calculations with U below 3.3 eV results on a shallow acceptor level, while calculations with U = 5.0 eV or higher leads to a localized level. The results of U = 3.3 eV or below explain the observed high electronic conductivity of Mg-doped LiCoO$_2$. However, these U values are not typically employed to study defects and impurities in LiCoO$_2$ or similar materials. Instead, self-consistently determined\cite{zhou_first-principles_2004}  U have been used.\cite{tatsumi_local_2008,koyama_defect_2012} For Co, the self-consistent U values range from 4.91 to  5.62 eV.\cite{zhou_first-principles_2004} As our results show, U values in this range lead to deep levels in Mg-doped LiCoO$_2$, which is inconsistent with the observed high electronic conductivity of this material. 

\section{Conclusions}

In the present work, we have studied the doping of LiCoO$_2$ with Mg employing the GGA and GGA+U methods. In particular, we explored the effect of the U parameter on the energetic, geometric and electronic properties of Mg in LiCoO$_2$.  Our results show that these properties for Mg located on a Co or a Li site in LiCoO$_2$ depend on the chosen U value. A similar dependence on U was also found in the properties of Li vacancy and Li interstitial defects in LiCoO$_2$. The similarity arises because the substitutional Mg and the Li defects lead to impurity states in LiCoO$_2$ with changes on the valency of Co in LiCoO$_2$. The strong dependence on U of the energetic, geometric and electronic properties is a direct consequence of the valency change of Co. Increasing the U value eventually changes the impurity states from shallow to deep levels. Conversely, if Mg on Co or Li sites forms complexes and no impurity states are introduced in LiCoO$_2$, the properties of such defect complexes are insensitive to the U value.

These results indicate that GGA/GGA+U methods may be used to study isovalent substitution in LiCoO$_2$. For aliovalent substitution, such as Mg on a Co or a Li site, the usefulness of GGA/GGA+U methods is limited because experimental or theoretical data from accurate \textit{ab initio} methods\cite{booth_towards_2013,shulenburger_quantum_2013,wagner_quantum_2014,wagner_effect_2014,foyevtsova_ab_2014} are needed to validate the U dependent results. Moreover, even if such data were available, one is left with the problem of a single U value not giving a reasonable overall description of the properties of LiCoO$_2$. For example, U values close to 3\cite{ensling_electronic_2010} and 5\cite{ensling_electronic_2010,juhin_angular_2010,koyama_defect_2012,xiong_atomic_2012} eV result in band gaps similar to experimental values, U = 3.3\cite{kramer_tailoring_2009} eV correctly describes some of the major features in the phase diagram of LiCoO$_2$ while U values close to 5\cite{zhou_first-principles_2004,chevrier_hybrid_2010} eV are needed to reproduce the measured average Li-intercalation potential. Our present results show that U values of 3.3 eV and lower can describe the high electronic conductivity of Mg-doped LiCoO$_2$ but U as low as 1.5 eV are needed to describe the solubility of Mg in LiCoO$_2$. Finally, even if a perfect U value could be found, it would not improve the oxygen molecule formation energy or the thermodynamic properties of materials without d electrons. Empirical corrections would still be required.

\begin{acknowledgments}
The work was supported by the Materials Sciences \& Engineering Division of the Office of Basic Energy Sciences, U.S. Department of Energy (DOE).
Research by PRCK was conducted at the Center for Nanophase Materials Sciences, which is sponsored at Oak Ridge National Laboratory by the Scientific User Facilities Division, Office of Basic Energy Sciences, U.S. Department of Energy.
\end{acknowledgments}

\providecommand{\noopsort}[1]{}\providecommand{\singleletter}[1]{#1}%
%

\end{document}